\documentclass[twocolumn,showpacs,eqsecnum,prb,amsmath,amssymb]{revtex4}
\usepackage[latin1]{inputenc}
\usepackage{graphicx}
\usepackage{dcolumn}
\usepackage{bm}

\begin{document}

\title{Breakdown of the Fermi-liquid regime in the 2D Hubbard model from a two-loop field-theoretical renormalization group approach}
\author{Hermann Freire}
\email{H.Freire@fkf.mpg.de} \affiliation{Max-Planck-Institute for
Solid State Research, D-70569 Stuttgart, Germany}
\author{Eberth Correa}
\author{Alvaro Ferraz}
 \affiliation{International Center for Condensed Matter Physics, Universidade de Brasilia, Caixa Postal
 04667, 70910-900 Brasilia-DF, Brazil}

\begin{abstract}
We analyze the particle-hole symmetric two-dimensional Hubbard model
on a square lattice starting from weak-to-moderate couplings by means of the field-theoretical
renormalization group (RG) approach up to two-loop order. This
method is essential in order to evaluate the effect of the
momentum-resolved anomalous dimension $\eta(\textbf{p})$ which
arises in the normal phase of this model on the corresponding
low-energy single-particle excitations. As a result, we find
important indications pointing to the existence of a non-Fermi
liquid (NFL) regime at temperature $T\rightarrow 0$ displaying a truncated Fermi surface
(FS) for a doping range exactly in between the well-known
antiferromagnetic insulating and the $d_{x^2-y^2}$-wave singlet
superconducting phases. This NFL evolves as a function of doping
into a correlated metal with a large FS before the
$d_{x^2-y^2}$-wave pairing susceptibility finally produces the
dominant instability in the low-energy limit.
\end{abstract}


\maketitle

\section{Introduction}

The physical nature of underdoped cuprates both above and below the
superconducting temperature $T_{c}$ is still subjected to strong
debate \cite{Millis,Timusk,Norman}. The recent experiment of Doiron-Leyraud
\textit{et al.} \cite{Leyraud} applying a magnetic field strong
enough to destroy the superconducting state established the
existence of quantum oscillations and coherent Fermi surface (FS)
pockets for this regime. By contrast, earlier angle-resolved
photoemission experiments (ARPES) reported only observations of
disconnected Fermi arcs located in the nodal regions of momentum
space \cite{Ding,Loeser}. Moreover, according to both scenarios, there
exist charge pseudogaps and no quasiparticle-like excitations in the
corresponding antinodal sectors centered around $(\pm\pi,0)$ and
$(0,\pm\pi)$. The absence of these quasiparticle excitations is
signalled by the vanishing of the quasiparticle peak $Z(\mathbf{p})$
in the spectral function at these antinodal regions. The pseudogap
behavior also reflects itself in the underdoped superconducting
phase. One manifestation of this is the fact that the
superconducting gap and the pseudogap behave distinctively as a
function of doping: the superconducting gap located in the nodal
regions centered around $(\pm\pi/2,\pm\pi/2)$ decreases continuously
whereas the pseudogap continues to increase as the hole doping is
reduced from its optimal value \cite{Shen,Tanaka}. Such a nodal-antinodal
dichotomy is therefore a common trend in underdoped cuprates.

A minimal model for describing the dynamics of the interacting
electrons in these high-$T_{c}$ superconductors is the single-band
2D Hubbard model \cite{Anderson}. This model has been investigated
extensively with several numerical techniques such as exact
diagonalization, quantum Monte Carlo and the more recent quantum
clusters approaches \cite{Maier,Kyung}, as well as by
semi-analytical functional renormalization group (RG) methods
\cite{Zanchi,Metzner,Honerkamp,Shankar,Kopietz,Tsai,Gonzalez,Binz,Ferraz}.
Several of these works successfully obtain an antiferromagnetic
phase in the model near half-filling and the onset of a
$d_{x^2-y^2}$-wave singlet superconducting phase away from
half-filling when the temperature is lowered below a critical value
\cite{Zanchi,Metzner,Honerkamp,Maier}. These results therefore give
further support to the point of view that the 2D Hubbard model might
indeed capture the essential aspects of the physics displayed by
those strongly-correlated materials.

On the experimental side, the insulating antiferromagnetic phase in
the hole-doped cuprates is quickly destroyed at a very small but
nevertheless nonzero doping. If we approach this phase from a larger
doping regime, this suggests that the existing Fermi surface should
shrink to zero with the corresponding quasiparticle weight
$Z(\mathbf{p})$ becoming suppressed all along the underlying FS at a
possible quantum critical point. It is therefore clearly important
to account for the physical effects associated with such a vanishing
of $Z$ at a nonzero doping near half-filling. This analysis has been
initiated in recent years for the 2D Hubbard model and its
extensions in the context of the fermionic RG framework
\cite{Zanchi2,Honerkamp2,Katanin,Metzner2}. However, since the
self-energy effects manifest themselves only at two-loop order or
beyond, in order to devise conserving many-body approximations \cite{comment} for
this model, one should necessarily evaluate all flow equations up to
the same order of perturbation theory. For this
reason, we implement in this work a two-loop field-theoretical RG
calculation as a function of doping concentration for some important
quantities in the 2D Hubbard model.

Following previous one-loop RG calculations performed in the 2D
Hubbard model, we choose to consider here only weak-to-moderate
initial couplings. As a result, we find important indications
pointing to the existence of a non-Fermi liquid (NFL) regime at a
nonzero doping, before the $d_{x^2-y^2}$-wave singlet superconducting
instability finally dominates over the antiferromagnetic
fluctuations. An important point we wish to stress here is that this
NFL regime emerges from the nonzero momentum-resolved anomalous
dimension $\eta(\mathbf{p})$ contribution to the single-particle
excitations, which arises only at a two-loop RG level or beyond
close to half-filling. In the corresponding one-loop RG flow
equations, even if the initial values are taken to be reasonably
small, the renormalized interactions rapidly flow towards a strong
coupling regime in the low-energy limit. This fact also indicates
the importance of higher-order quantum corrections to the full
description of the low-energy dynamics of the 2D Hubbard model. Our
work therefore takes seriously this observation and, for this
reason, it represents a step forward in this direction.

This paper is organized as follows. In Sec. II, we explain the
methodology employed to discuss the 2D Hubbard model. In this part,
we will choose to be very schematic since the fermionic
field-theoretical RG methodology was already explained at length and
in full detail in the context of a simpler 2D flat FS model
elsewhere \cite{Hermann}. Our main emphasis will be rather to
highlight the final integro-differential two-loop RG flow equations
resulting from the application of this method to the 2D Hubbard
model. In Sec. III, we move on to the numerical solution of these RG
flow equations. Lastly, in Sec. IV, we present our final conclusions
regarding our two-loop RG calculation and we point out open issues that
still have to addressed for the full clarification of this important
problem.

\begin{figure}[t]
  \includegraphics[width=2.35in]{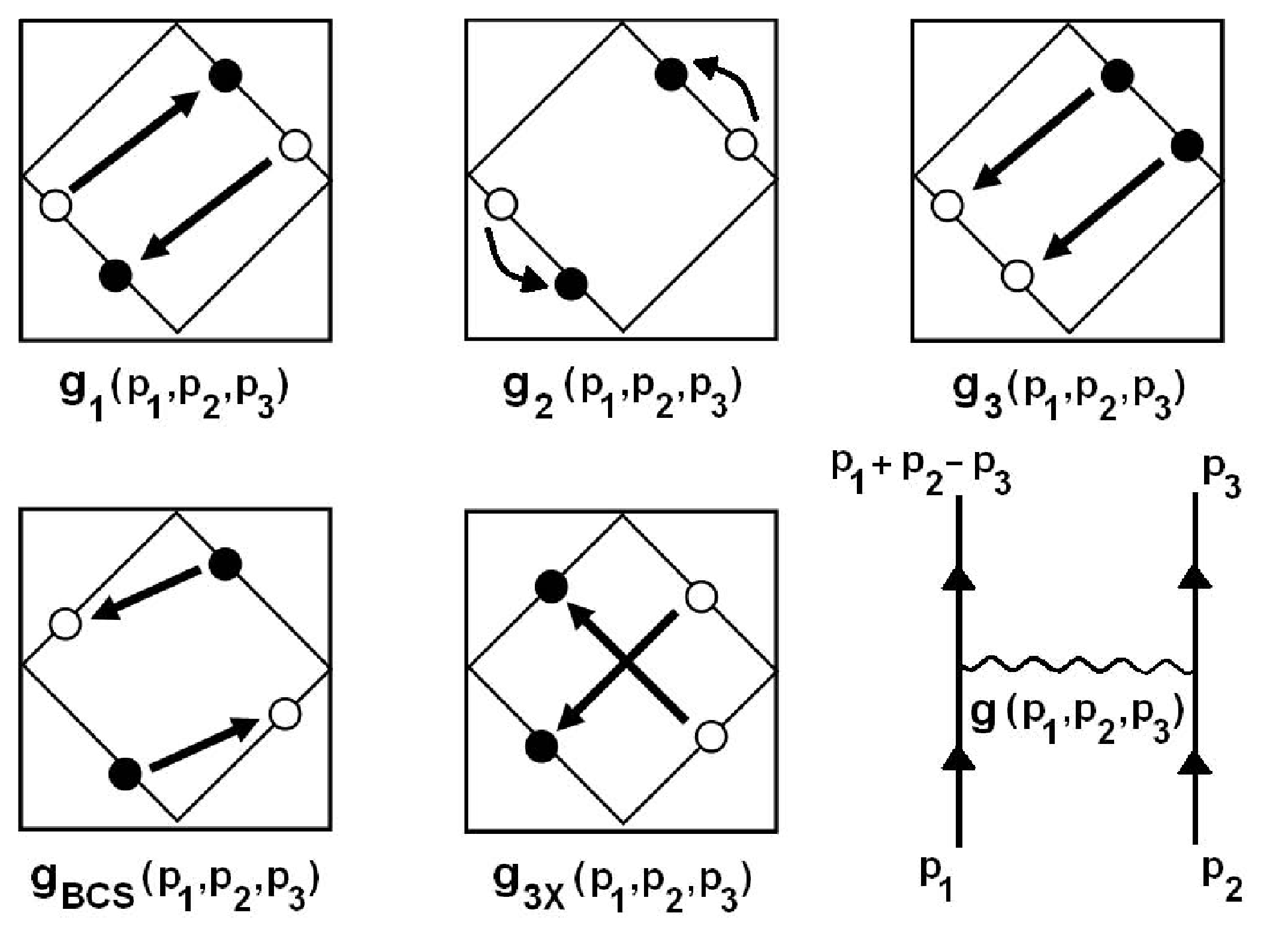}\\
  \caption{The g-ology parametrization for the marginally relevant interaction processes of the
  2D Hubbard model, using as a reference the
corresponding noninteracting FS at half-filling for
simplicity.}\label{A}
\end{figure}

\section{Methodology}

We start by defining the Hamiltonian of the 2D
Hubbard model in momentum space

\vspace {-0.2cm}

\begin{equation}
H=\sum_{\mathbf{k},\sigma}\xi_{\mathbf{k}}\psi^{\dagger}_{\mathbf{k}\sigma}\psi_{\mathbf{k}\sigma}+\left(\frac{U}{N_{s}}\right)
\sum_{\mathbf{p},\mathbf{k},\mathbf{q}}\psi^{\dagger}_{\mathbf{p}+\mathbf{k}-\mathbf{q}\uparrow}\psi^{\dagger}_{\mathbf{q}\downarrow}
\psi_{\mathbf{k}\downarrow}\psi_{\mathbf{p}\uparrow}.
\end{equation}

\noindent where $\psi^{\dagger}_{\mathbf{k}\sigma}$ and
$\psi_{\mathbf{k}\sigma}$ are the usual fermionic creation and
annihilation operators with momentum $\mathbf{k}$ and spin
projection $\sigma$, $t$ is the electronic hopping amplitude to
nearest neighbor sites, $\mu$ is the chemical potential which
controls the band filling (and, consequently, the doping parameter),
$U$ is the local on-site repulsive interaction and $N_{s}$ is the
total number of lattice sites. Since we will be interested only in the
universal quantities of this model, we can linearize the
tight-binding energy dispersion
$\xi_{\mathbf{k}}=-2t\big[\cos(k_{x})+\cos(k_{y})\big]-\mu$ around
the FS as $\xi_{\mathbf{k}}\approx
v_{F}(\mathbf{k})\mathbf{\hat{n}}.\big(\mathbf{k-k_{F}}(\mu)\big)$
with the Fermi velocity given by
$v_{F}(\mathbf{k})=\big|(\nabla_{\mathbf{k}}\xi_{\mathbf{k}}\small|_{\mathbf{k=k_{F}}(\mu)})\big|=2t\sqrt{\sin^{2}k_{x}+\sin^{2}k_{y}}$,
$\mathbf{k_{F}}(\mu)$ being the Fermi momentum which defines the
noninteracting FS for a continuous doping parameter and
$\mathbf{\hat{n}}$ is a unit vector perpendicular to the FS.

The thermodynamical properties of this model can be computed from
the coherent-state Grassmann representation of the partition
function

\vspace{-0.2cm}

\begin{equation}
Z=\int \mathcal{D}[\bar{\psi},\psi]e^{-\int_{0}^{\beta} d\tau(L_{0}[\bar{\psi},\psi]+L_{int}[\bar{\psi},\psi])},
\end{equation}

\noindent where $\beta=1/T$. The noninteracting Lagrangian is defined in a standard way

\vspace{-0.2cm}

\begin{equation}
L_{0}[\bar{\psi},\psi]=\sum_{\sigma}\int_{\mathbf{k}} \bar{\psi}_{\sigma}(\mathbf{k},\tau)\big(\partial_{\tau}+\xi_{\mathbf{k}}\big)\psi_{\sigma}(\mathbf{k},\tau),\label{2}
\end{equation}

\noindent where $\int_{\mathbf{k}}=\int\frac{d^{2}\mathbf{k}}{(2\pi)^2}$ and the interacting Lagrangian in turn reads

\vspace{-0.2cm}

\begin{eqnarray}
&&L_{int}[\bar{\psi},\psi]=\sum_{\sigma,\sigma'}\int_{\mathbf{p_{1}}}\int_{\mathbf{p_{2}}}\int_{\mathbf{p_{3}}}g(\mathbf{p_{1},p_{2},p_{3}})\nonumber\\
&&\times\bar{\psi}_{\sigma}(\mathbf{p_{1}+p_{2}-p_{3}},\tau)\bar{\psi}_{\sigma'}(\mathbf{p_{3}},\tau)\psi_{\sigma'}(\mathbf{p_{2}},\tau)
\psi_{\sigma}(\mathbf{p_{1}},\tau).\nonumber\\
\label{3}
\end{eqnarray}

\noindent The Eqs. (\ref{2}) and (\ref{3}) therefore define our bare
quantum field theory which is regularized in the ultraviolet by
restricting the momenta to $|\mathbf{k}|\leq \Lambda_{0}$, where the
cutoff is chosen to be $\Lambda_{0}=4t$. Since this should
correspond originally to the 2D Hubbard model, we must set the bare
interaction $g(\mathbf{p_{1},p_{2},p_{3}})$ initially equal to the
local interaction $U$. However, as we will see next, this functional
dependence of the coupling on the momenta will in fact play a
crucial role in the low-energy effective theory.

Here we perform all the calculations in the $T\rightarrow0$ limit.
The starting point of our approach is the noninteracting FS of the
2D Hubbard model defined by $\mathbf{k_{F}}(\mu)$ for a continuous
doping parameter. To include the effect of interactions, we use a
g-ology parametrization adapted appropriately to our 2D problem (for
more details on this procedure in the context of a simpler 2D flat
FS model, see, e.g., Ref. \cite{Hermann}). By means of a power
counting analysis, one can easily verify that the dependence of the
coupling function $g(\mathbf{p_{1},p_{2},p_{3}})$ on the components
of the momenta normal to the FS are irrelevant in the RG sense.
Therefore, we are allowed to project these coupling functions on the
FS in such a way that their only functional dependence will come
from the components parallel to the FS of the three external momenta
$\mathbf{p_{1}}$, $\mathbf{p_{2}}$, and $\mathbf{p_{3}}$ (the fourth
momentum is given naturally by momentum conservation
$\mathbf{p_{4}}=\mathbf{p_{1}}+\mathbf{p_{2}}-\mathbf{p_{3}}$). In
this way, we consider here the interaction processes which lead to
singularities within perturbation theory in the low-energy limit.
They are shown schematically in Fig. 1. These processes are the
marginally relevant couplings in our RG theory, since their
contributions become increasingly important at low energies. We
neglect here the so-called $g_{4}$-processes of interacting
particles belonging to the same FS sector. Experience with
one-dimensional systems suggests that they should not alter
qualitatively our results. Besides, we also neglect all marginally
irrelevant interactions, since these contributions are not expected
to change the universal properties of the model.

The methodology of our RG scheme follows closely the
field-theoretical method \cite{Peskin,Weinberg}. In perturbation theory, infrared
logarithmic singularities typically emerge in the low-energy
limit at the calculation of several quantities in the model such
as the vertex corrections, the quasiparticle weight and the
susceptibilities. We circumvent this
problem by introducing appropriate counterterms at a flowing RG scale
parameter $\Lambda$, in such a way that all the observables -- i.e.
the renormalized quantities of the theory -- remain well-defined in
the low-energy limit ($\Lambda\rightarrow 0$). As we have already
pointed out in an earlier paper \cite{Hermann}, the problem of
formulating a RG theory associated with a FS of a 2D model often
requires the definition of counterterms which are continuous
functions of the parallel momenta along the FS. In this way, we must
perform at two-loop level the following substitutions for the
fermionic fields

\vspace {-0.2cm}

\begin{eqnarray}
&&\psi_{\sigma}(\mathbf{p},\tau)\rightarrow Z_{\Lambda}^{1/2}(\mathbf{p})\psi_{\sigma}(\mathbf{p},\tau),\nonumber\\
&&\bar{\psi}_{\sigma}(\mathbf{p},\tau)\rightarrow Z_{\Lambda}^{1/2}(\mathbf{p})\bar{\psi}_{\sigma}(\mathbf{p},\tau),
\label{definition}
\end{eqnarray}

\noindent where, from now on, all the momenta will correspond to the
momentum projected on the FS. Besides, $Z_{\Lambda}(\mathbf{p})$ is
the RG flowing momentum-resolved quasiparticle weight and it is
naturally related in the limit of $\Lambda\rightarrow 0$ to the
conventional many-body definition of the quasiparticle peak
$Z(\mathbf{p})=(1-\partial\rm
Re\Sigma(\omega,\mathbf{p})/\partial\omega|_{\omega=0})^{-1}$. Using
Eq. (\ref{definition}), we also find that the bare and renormalized
coupling functions are related to each order by

\vspace {-0.2cm}

\begin{eqnarray}
&&g_{i}(\mathbf{p_{1}},\mathbf{p_{2}},\mathbf{p_{3}})=\left[\prod_{j=1}^{4}Z_{\Lambda}^{-1/2}(\mathbf{p_{j}})\right]
\bigg(g_{iR}(\mathbf{p_{1}},\mathbf{p_{2}},\mathbf{p_{3}};\Lambda)\nonumber\\
&&+\Delta g_{iR}^{1loop}(\mathbf{p_{1}},\mathbf{p_{2}},\mathbf{p_{3}};\Lambda)+\Delta g_{iR}^{2loops}(\mathbf{p_{1}},\mathbf{p_{2}},\mathbf{p_{3}};\Lambda)\bigg),\nonumber\\
\label{gbare}
\end{eqnarray}

\noindent where $i=$1, 2, 3, 3X, and BCS. The renormalized
quantities (labeled by the subscript $R$) generally depend on the RG
scale $\Lambda$. In contrast, all the bare quantities will be denoted here without any additional
index. The functions $\Delta
g_{iR}^{1loop}(\mathbf{p_{1}},\mathbf{p_{2}},\mathbf{p_{3}};\Lambda)$
and $\Delta
g_{iR}^{2loops}(\mathbf{p_{1}},\mathbf{p_{2}},\mathbf{p_{3}};\Lambda)$
represent the counterterms necessary to regularize at one-loop and
two-loops respectively the four-point one-particle irreducible (1PI) vertex corrections in each
of the corresponding parametrized scattering channels.

\begin{figure}[t]
  \includegraphics[width=3.4in,]{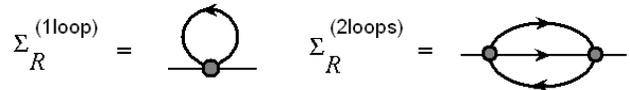}
  \\
  \caption{The self-energy diagrams up to two-loop order. The one-loop term is the
Hartree diagram and the two-loop contribution is the so-called sunset diagram.} \label{B}
\end{figure}

Following the same strategy as it was described in full detail in
Ref. \cite{Hermann}, we now calculate the quasiparticle weight
$Z_{\Lambda}(\mathbf{p})$ at two-loop level starting from the
definition of the renormalized self-energy of the present model. The
corresponding Feynman diagrams up to two-loop order are shown in
Fig. 2. The one-loop diagram (i.e., the Hartree term) is generally
independent of the external frequency $\omega$. As a result, this
contribution never renormalizes the quasiparticle weight. In fact,
it only generates a constant shift in the chemical potential of the
model which must be appropriately subtracted by a counterterm in
such a way that the density of particles in the system always
remains fixed during the RG flow. By contrast, the two-loop
contribution (i.e., the sunset diagram) is the first contribution to
the self-energy which produces a nonanalyticity as a function of the
external frequency $\omega$. For this reason, this term alone will
be responsible for the renormalization of the quasiparticle weight
$Z_{\Lambda}(\mathbf{p})$ at this level of perturbation theory as we
vary the RG scale $\Lambda$ towards the low-energy limit.

The momentum-resolved anomalous dimension is conventionally defined
by $\eta(\mathbf{p})=\Lambda d\ln Z_{\Lambda}(\mathbf{p})/d\Lambda$.
From this expression, we get

\vspace {-0.2cm}

\begin{eqnarray}
\eta(\mathbf{p})&=&\frac{1}{8\pi^{4}}\int_{FS} d\mathbf{k}\int_{FS} d\mathbf{q} \left[\frac{\Lambda}{\Lambda+2|\mu|F(\mathbf{p},\mathbf{k},\mathbf{q})}\right]\nonumber\\ &\times&\big[2g_{1R}(\mathbf{p},\mathbf{k+q-p},\mathbf{k})g_{1R}(\mathbf{k},\mathbf{q},\mathbf{p})\nonumber\\
&+&2g_{2R}(\mathbf{p},\mathbf{k}
\mathbf{+q-p},\mathbf{q})g_{2R}(\mathbf{k},\mathbf{q},\mathbf{k+q-p})
\nonumber\\&-&g_{1R}(\mathbf{p},\mathbf{k+q-p},\mathbf{k})
g_{2R}(\mathbf{k},\mathbf{q},\mathbf{k+q-p})\nonumber\\
&-&g_{2R}(\mathbf{p},\mathbf{k+q-p},\mathbf{q})g_{1R}(\mathbf{k},\mathbf{q},\mathbf{p})+2g_{3R}(\mathbf{k},\mathbf{p},\mathbf{q})\nonumber\\
&\times& g_{3R}(\mathbf{k},\mathbf{p},\mathbf{q})-g_{3R}(\mathbf{k},\mathbf{p},\mathbf{q})g_{3R}(\mathbf{p},\mathbf{k},\mathbf{q})\big]\nonumber\\
&\times&\left(\frac{1}{v_{F}(\mathbf{k})+v_{F}(\mathbf{q})}\right)\left(\frac{1}{v_{F}(\mathbf{q})+v_{F}(\mathbf{k+q-p})}\right),\nonumber\\
\nonumber\\
\label{eta}
\end{eqnarray}

\noindent where the integrals over the momenta now are simply along the curve defined by the FS and, besides, $F(\mathbf{p},\mathbf{k},\mathbf{q})=v_{F}(\mathbf{k+q-p})/v_{F}(\mathbf{k})$. It is interesting to note here that if
we take the 1D limit of the above equation, we reproduce exactly the
well-known result for the anomalous dimension of the Luttinger
liquid at two-loop order as was first calculated long ago in Ref. \cite{Solyom}.

To derive the two-loop RG flow equations for the renormalized coupling functions, we must take into account
the fact that the bare theory does not know anything about the RG scale $\Lambda$. In other words,
the bare couplings do not depend on this scale, i.e. $\Lambda dg_{i}(\mathbf{p_{1}},\mathbf{p_{2}},\mathbf{p_{3}})/d\Lambda=0$. Therefore,
using Eq. (\ref{gbare}), we finally obtain

\vspace {-0.2cm}

\begin{eqnarray}
&&\Lambda\frac{d}{d\Lambda}g_{iR}(\mathbf{p_{1}},\mathbf{p_{2}},\mathbf{p_{3}})=\frac{1}{2}\sum_{i=1}^{4}\eta(\mathbf{p_{i}})
g_{iR}(\mathbf{p_{1}},\mathbf{p_{2}},\mathbf{p_{3}})\nonumber\\
&&-\Lambda\frac{d}{d\Lambda}\Delta
g_{iR}^{1loop}(\mathbf{p_{1}},\mathbf{p_{2}},\mathbf{p_{3}})
-\Lambda\frac{d}{d\Lambda}\Delta g_{iR}^{2loops} (\mathbf{p_{1}},\mathbf{p_{2}},\mathbf{p_{3}}),\nonumber\\
\label{RGequations}
\end{eqnarray}

\noindent where $i=$1, 2, 3, 3X, and BCS. The Feynman diagrams for
the vertex corrections necessary for the determination of the
counterterm functions are shown schematically in Fig. 3. These RG
flow equations for the renormalized couplings are in fact
complicated integro-differential equations coupled to one another.
We will discuss their numerical solution in the next section.

\begin{figure}[t]
  \includegraphics[width=3.2in,]{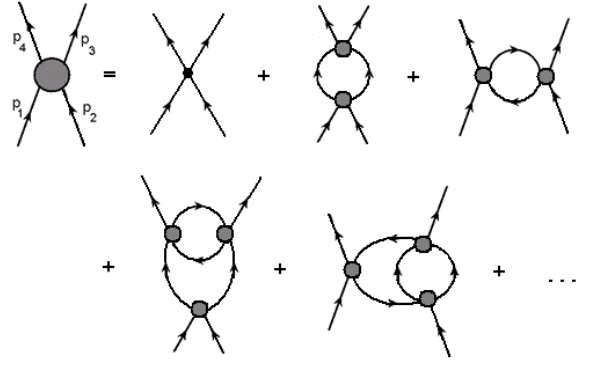}
  \\
  \caption{Some Feynman diagrams showing the vertex corrections up to two-loop order.} \label{B}
\end{figure}

To investigate what are the enhanced correlations in low-energy
limit of the 2D Hubbard model, we must calculate the linear response
of the system to small external fields. Therefore, we must add to
our original Lagrangian the folllowing term

\vspace {-0.2cm}

\begin{eqnarray}
L_{ext}&=&\int_{\mathbf{q,p}}\bigg[h_{SC}(\mathbf{q})\gamma^{\alpha\beta}_{SC}(\mathbf{p,q})\bar{\psi}_{\alpha}
(\mathbf{p},\tau)\bar{\psi}_{\beta}(\mathbf{-p+q},\tau)\nonumber\\
&+&h_{DW}(\mathbf{q})\gamma^{\alpha\beta}_{DW}(\mathbf{p,q})\bar{\psi}_{\alpha}
(\mathbf{p+q},\tau)\psi_{\beta}(\mathbf{p},\tau)+h.c.\bigg]\nonumber\\
\end{eqnarray}

\noindent where $h_{SC}(\mathbf{q})$ and $h_{DW}(\mathbf{q})$ are
the external fields and $\gamma^{\alpha\beta}_{SC}(\mathbf{p,q})$
and $\gamma^{\alpha\beta}_{DW}(\mathbf{p,q})$ are the response
vertices for superconducting and density-wave orders respectively \cite{Eberth}.
This added term will generate new Feynman diagrams -- the
three-legged vertices displayed in Fig. 4 -- which will also
generate new logarithmic singularities in the low-energy limit of
our quantum field theory. Therefore, we must regularize these
divergences by defining new counterterms as follows

\vspace {-0.3cm}

\begin{eqnarray}
\gamma^{\alpha\beta}_{SC}(\mathbf{p,q})&=&Z_{\Lambda}^{-1/2}(\mathbf{p})Z_{\Lambda}^{-1/2}(\mathbf{-p+q})\bigg[\gamma^{\alpha\beta}_{R,SC}(\mathbf{p,q};\Lambda)\nonumber\\
&+&\Delta\gamma^{\alpha\beta}_{R,SC}(\mathbf{p,q};\Lambda)\bigg]\nonumber\\
\gamma^{\alpha\beta}_{DW}(\mathbf{p,q})&=&Z_{\Lambda}^{-1/2}(\mathbf{p+q})Z_{\Lambda}^{-1/2}(\mathbf{p})\bigg[\gamma^{\alpha\beta}_{R,DW}(\mathbf{p,q};\Lambda)
\nonumber\\
&+&\Delta\gamma^{\alpha\beta}_{DW}(\mathbf{p,q};\Lambda)\bigg]
\end{eqnarray}

\noindent where the $Z$-factors, as before, come from the
redefinition of the fermionic fields at two-loop RG level displayed
in Eq. (\ref{definition}). As a result, using again the
fact that the bare response vertices do not depend on the RG scale
$\Lambda$, we finally get

\vspace {-0.2cm}

\begin{eqnarray}
\Lambda\frac{d}{d\Lambda}\gamma^{\alpha\beta}_{R,SC}(\mathbf{p,q})&=&\frac{1}{2}\big[\eta(\mathbf{p})+\eta(\mathbf{-p+q})\big]\gamma^{\alpha\beta}_{R,SC}(\mathbf{p,q})\nonumber\\
&-&\Lambda\frac{d}{d\Lambda}\Delta\gamma^{\alpha\beta}_{R,SC}(\mathbf{p,q})\nonumber\\
\Lambda\frac{d}{d\Lambda}\gamma^{\alpha\beta}_{R,DW}(\mathbf{p,q})&=&\frac{1}{2}\big[\eta(\mathbf{p+q})+\eta(\mathbf{p})\big]\gamma^{\alpha\beta}_{R,DW}(\mathbf{p,q})\nonumber\\
&-&\Lambda\frac{d}{d\Lambda}\Delta\gamma^{\alpha\beta}_{R,DW}(\mathbf{p,q})
\label{RGequations2}
\end{eqnarray}

\begin{figure}[b]
  \includegraphics[width=3.2in]{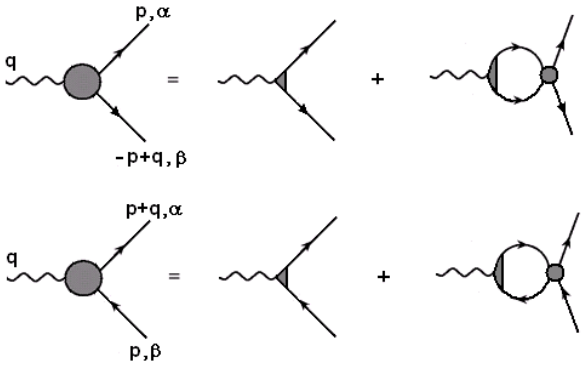}
  \\
  \caption{The Feynman diagrams for the three-legged response vertices associated with
superconducting and density-wave orders.} \label{B}
\end{figure}

\noindent Now, if we symmetrize the response vertices $\gamma^{\alpha\beta}_{SC}(\mathbf{p,q})$
and $\gamma^{\alpha\beta}_{DW}(\mathbf{p,q})$ with respect to their spin indices we obtain the following
order parameters

\vspace {-0.2cm}

\begin{eqnarray}
&&\gamma_{SSC}(\mathbf{p,q})=\gamma^{\uparrow\downarrow}_{SC}(\mathbf{p,q})-\gamma^{\downarrow\uparrow}_{SC}(\mathbf{p,q})\nonumber\\
&&\gamma_{TSC}(\mathbf{p,q})=\gamma^{\uparrow\downarrow}_{SC}(\mathbf{p,q})+\gamma^{\downarrow\uparrow}_{SC}(\mathbf{p,q})\nonumber\\
&&\gamma_{SDW}(\mathbf{p,q})=\gamma^{\uparrow\uparrow}_{DW}(\mathbf{p,q})-\gamma^{\downarrow\downarrow}_{DW}(\mathbf{p,q})\nonumber\\
&&\gamma_{CDW}(\mathbf{p,q})=\gamma^{\uparrow\uparrow}_{DW}(\mathbf{p,q})+\gamma^{\downarrow\downarrow}_{DW}(\mathbf{p,q})\nonumber\\
\end{eqnarray}

\noindent where SSC and TSC correspond to singlet and triplet superconductivity and SDW and CDW stand for charge and spin
density waves, respectively. Hence, using the above relations one can readily derive the RG flow equations for
each response vertex associated with a potential instability of the normal state towards a given ordered (i.e. symmetry-broken) phase.
The initial conditions for these RG flow equations will determine the symmetry of the order parameter. Thus

\begin{eqnarray}
\gamma_{iR}(\mathbf{p,q};\Lambda=\Lambda_{0})&=&1 \hspace{0.2cm} \textrm{($s$-wave)}\nonumber\\
\gamma_{iR}(\mathbf{p,q};\Lambda=\Lambda_{0})&=&\frac{1}{\sqrt{2}}(\cos p_{x}-\cos p_{y}) \hspace{0.2cm} \textrm{($d_{x^{2}-y^{2}}$-wave)}\nonumber\\
\end{eqnarray}

\noindent where $i=$ SSC, TSC, CDW and SDW. Once we computed the response vertices associated with these order parameters, we can now proceed to calculate their corresponding static susceptibilities. From Fig. 5, one can easily verify that they are given by

\begin{eqnarray}
&&\Lambda \frac{d}{d\Lambda}\chi_{R,SC}(\mathbf{q}=0;\Lambda)=-\int \frac{d\mathbf{p}}{v_{F}(\mathbf{p})}  [\gamma_{R,SC}(\mathbf{p,-p};\Lambda)]^{2}\nonumber\\
&&\Lambda \frac{d}{d\Lambda}\chi_{R,DW}(\mathbf{q}=\mathbf{Q};\Lambda)=-\left(\frac{\Lambda}{\Lambda+2|\mu|}\right)\int \frac{d\mathbf{p}}{v_{F}(\mathbf{p})} \nonumber\\
&&\hspace{4cm} \times[\gamma_{R,DW}(\mathbf{p+Q,p};\Lambda)]^{2}\nonumber\\
\end{eqnarray}

\noindent where, to avoid cluttering up the notation too much, we simply write $SC$ for both singlet and triplet and $DW$ for charge and spin density wave susceptibilities. In addition, the vector $\mathbf{Q}=(\pi,\pi)$ is the commensurate antiferromagnetic wave vector.

\begin{figure}[t]
  \includegraphics[width=2.2in]{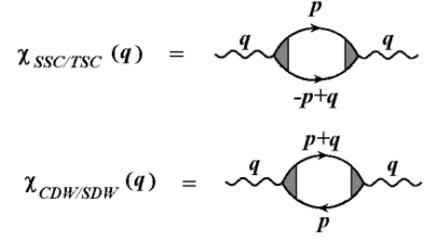}
  \\
  \caption{Feynman diagrams for the susceptibilities associated with both singlet and triplet superconductivity
and charge and spin density wave orders.} \label{B}
\end{figure}

\section{Numerical Results}

To solve all these integro-differential RG equations numerically, we will follow the standard
procedure of discretizing the full FS continuum into $N$ patches and then apply
the fourth-order Runge-Kutta method.
All numerical results in this paper will be presented for $N=32$.
Despite this moderate choice of number of patches, we point out that our results show good convergence properties
in the low-energy limit.

\begin{figure}[b]
  \includegraphics[width=2.4in, angle=270]{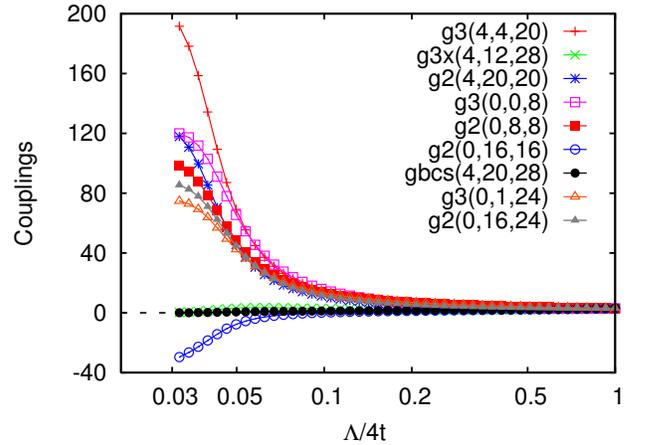}
  \\
  \caption{(Color online) The two-loop RG flow of the renormalized couplings (in units of $t$)
  for initial interaction $U=3t$ and $\mu=-0.05t$. We discretize the FS into $N=32$ patches, where 0, 8, 16, 24 represent
the antinodal directions, and 4, 12, 20, 28 refer to the nodal points.} \label{B}
\end{figure}

First, we focus our attention on the numerical solution of Eq.
(\ref{RGequations}) for the renormalized couplings as a function of
the ratio $\Lambda/4t$. As an initial condition for these equations,
we choose the weak-to-moderate bare interaction $U=3t$. Our results
are displayed in Fig. 6 for the case of $\mu=-0.05t$, which
corresponds to a hole doping $x$ of approximately $x\approx 3.4\%$.
In this regime, we observe that the inclusion of two-loop quantum
fluctuations has important consequences to the flow of all
couplings. Instead of exhibiting a divergence at a finite energy
scale as obtained in several earlier one-loop RG investigations
\cite{Zanchi,Metzner,Honerkamp}, the couplings now display a
tendency to level off in the low-energy limit. This result therefore
implies that the two-loop RG corrections clearly diminish the
importance of fluctuations as compared to the one-loop RG theory. It
is true, however, that several (but not all) couplings become
saturated at fairly strong-coupling fixed values. This is a
well-known problem in RG theory and happens as well in other
applications of this method to quantum field theories in which
fluctuation effects are known to be very strong (the most notorious
example being the Wilson-Fisher fixed point \cite{Wilson,Brezin} in
$\phi^{4}$-theory at three dimensions). Nevertheless, the RG results
in most cases are qualitatively correct even in a strong-coupling
regime. The same is true in our case. When one of the
renormalization couplings reaches an upper bound at the FS the whole
flow is automatically stopped in spite of the fact that all
couplings will not behave in the same way. We test the validity of
our results by analyzing several physical quantities simultaneously.
Until this critical RG scale is reached by one of the
couplings, no unphysical trend is observed in our calculations. In view of this, we hope that our two-loop RG results will also
capture at least qualitatively the most essential aspects of the
physics of the 2D Hubbard model in the low-energy limit.

\begin{figure}[t]
  \includegraphics[height=3.1in, angle=270]{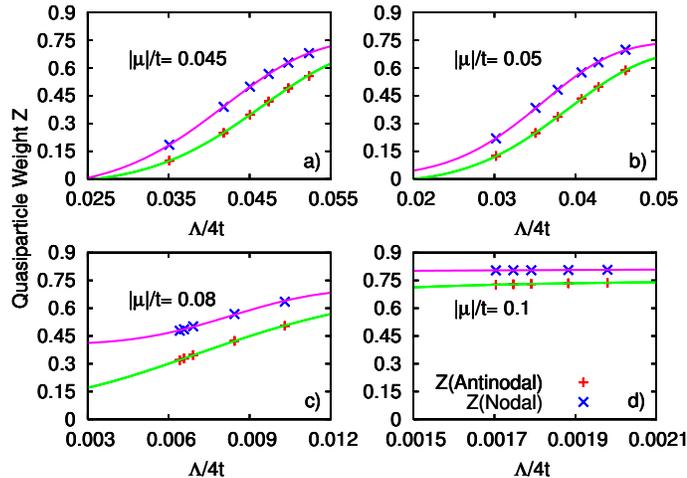}
  \\\vspace{0.1in}
  \caption{(Color online) The renormalized quasiparticle weight $Z_{\Lambda}(\mathbf{k})$ in the nodal and antinodal directions
versus $\Lambda/4t$ for initial interaction
  $U=3t$ and initial condition $Z_{\Lambda_{0}=4t}(\mathbf{k})=1$: a) $|\mu|/t=0.045$, b) $|\mu|/t=0.05$, c) $|\mu|/t=0.08$ and d) $|\mu|/t=0.1$.
The solid lines represent curve fittings of the RG
 data.}\label{C}
\end{figure}

Next we present the results for the two-loop RG flow of the
associated momentum-resolved quasiparticle weight
$Z_{\Lambda}(\textbf{p})$. Here $Z_{\Lambda}(\textbf{p})$ is
determined for several doping regimes as the low-energy limit is
approached. The RG flows are displayed in Fig. 7 for the special
choices of momentum at the nodal [near $(\pi/2,\pi/2)$] and
antinodal [near $(\pi,0)$] directions. For doping levels up to a
critical value $\mu_{c}=-0.045t$, corresponding to $x_{c}\approx
3\%$, we observe that the quasiparticle weights associated with both
nodal and antinodal directions become strongly suppressed and
eventually scale down to zero at sufficiently low-energy scales.
This result in fact holds for all points of the underlying FS of the
system. From the two-loop RG flows of the several order-parameter
susceptibilities displayed in Figs. 8(a) and 8(b), we observe that
the s-SDW is by far the dominant instability at these low doping
values. From these two results we can infer that, for
$|\mu|/t\leq0.045$, the Fermi surface (if we define it simply as the
locus in $\mathbf{k}$-space which exhibits a discontinuous jump in
the momentum distribution function)  is completely smeared out by
interactions and the resulting ground state should be given by an
antiferromagnetic insulator. This agrees for instance with the fact
that, exactly at half-filling, this model is known to have an
antiferromagnetic insulating ground state for any finite local
interaction $U$. Our two-loop RG approach therefore successfully
reproduces this important aspect of the problem.

\begin{figure}[b]
  \includegraphics[height=3.1in,angle=270]{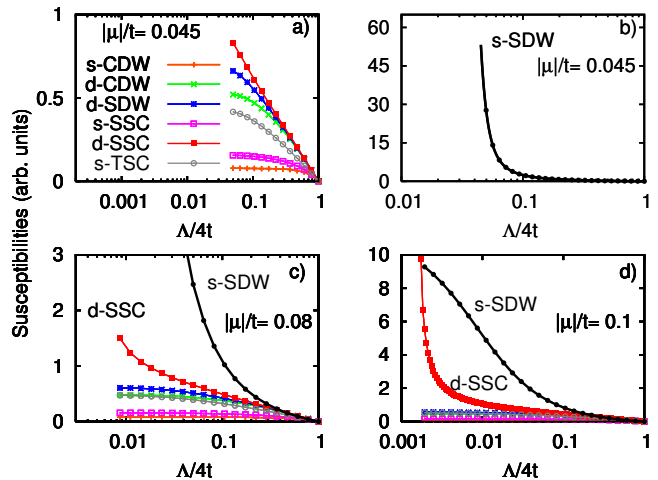}
  \\\vspace{0.1in}
  \caption{(Color online) The two-loop RG flows of several static order-parameter susceptibilities
  for initial interaction $U=3t$ with: a) and b) $|\mu|/t=0.045$, c) $|\mu|/t=0.08$ and d) $|\mu|/t=0.1$;
  CDW and SDW stand for charge and spin density waves; SSC and TSC stand for singlet and triplet superconductivity;
  the prefixes ``s" and ``d" refer to $s$-wave and $d_{x^2-y^2}$-wave symmetries.}\label{C}
\end{figure}

Following this, we probe the system at a slightly larger hole doping
regime, i.e. $\mu/t=-0.05$, which corresponds to $x\approx 3.4\%$.
At this regime, we obtain that the quasiparticle gap immediately
closes around the nodal directions but remains unaffected in the
antinodal regions. The complete suppression of the spectral weight
only in the antinodal zones suggests the existence of a NFL phase
with quasiparticle-like excitations only around the nodal directions
of the Fermi surface. From a technical point of view, this result
originates from the anisotropy in the anomalous dimension
$\eta(\mathbf{p})$ displayed in Eq. (\ref{eta}). This effect is to
some extent reminiscent of the nodal-antinodal dichotomy and is in
agreement with the general observation that, at light dopings, there
are no coherent quasiparticle peaks at the antinodal points in the
pseudogap phase of underdoped cuprates. However, we point out that
the anisotropy obtained here in this work is still very small as
compared with the experimental values observed in these materials.
This is undoubtedly related to our choice of initial interaction
$U=3t$. If we increase the on-site interaction of the model, we
expect that this trend will be amplified further. This conjecture
finds support in recent numerical results obtained from quantum
cluster approaches \cite{Kyung} for $U=8t$, where a stronger
nodal-antinodal dichotomy is indeed observed in the model at finite
doping.

Such a NFL character is observed here until the
corresponding doping reaches $|\mu|/t=0.08$ (i.e. $x\approx 5.1\%$).
At this hole doping, both the antinodal and nodal quasiparticle
weights become nonzero even when we extrapolate our RG flow to very
low $\Lambda/4t$ scales. This indicates that the resulting FS is
fully reconstructed in this regime. Moreover, as shown in Fig. 8(c),
we observe already the beginning of the unlimited growth of the
$d_{x^2-y2}$-wave pairing instability in the presence of the still
leading s-SDW susceptibility. This mixed d-SSC and s-SDW metallic
state is observed until $|\mu|/t=0.1$ (i.e. $x\approx 6.2\%$). At
this doping value, as indicated in Fig. 8(d), the d-SSC
susceptibility finally overcomes the s-SDW and the system turns into
a $d_{x^2-y^2}$-wave superconductor. As a consequence, we point out
that this latter result adds further support to the possible
existence of a $d_{x^2-y^2}$-wave superconducting state in the 2D
Hubbard model as was first obtained in Refs.
\cite{Zanchi,Metzner,Honerkamp} within a one-loop RG scheme.

One important point we would like to emphasize here is that, for doping regimes satisfying $|\mu|\leq 0.08t$, the
dominant fluctuations in our results are always of s-SDW type (i.e.
antiferromagnetism), in qualitative agreement with several previous
studies focusing on this model \cite{Maier,Zanchi,Metzner,Honerkamp}. This result therefore
favors the interpretation that the driving mechanism underlying our NFL evidence should be mediated
by antiferromagnetic spin-fluctuations.

\begin{figure}[t]
  \includegraphics[width=3.2in]{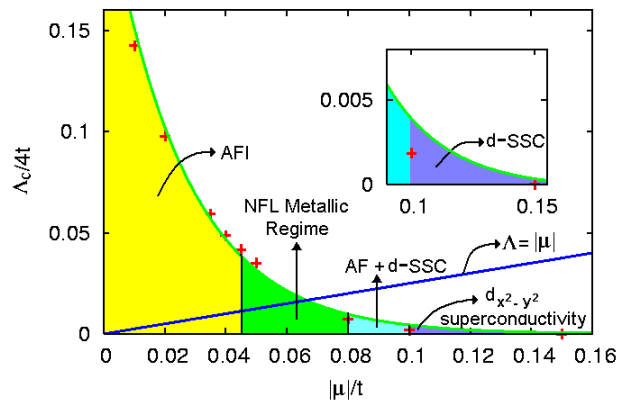}
  \\
  \caption{(Color online) Two-loop RG phase diagram for the 2D Hubbard model as a function of the critical RG scale
$\Lambda_{c}$ and the hole doping $\mu$ for local on-site bare interaction $U=3t$.}\label{C}
\end{figure}

We summarize our results displaying in Fig. 9 a two-loop RG phase
diagram of the 2D Hubbard model as a function of both a critical RG
scale $\Lambda_{c}$ and the hole doping represented by $\mu$. This
phase diagram however must be interpreted only qualitatively since
the numerical determination of the exact critical scale where the
order-parameter susceptibilities truly diverge is of course very
difficult and, for this reason, this is always based on a chosen
criterion. Therefore, as we saw before, we find evidence of an
extended antiferromagnetic insulating (AFI) phase from half-filling
up to $|\mu|/t=0.045$. For $0.045\leq|\mu|/t\leq0.08$, we obtain
instead that the formation of a NFL metallic phase with a truncated
Fermi surface in the nodal regions is favored. For
$0.08\leq|\mu|/t\leq0.1$, the FS is fully restored with the
resulting metallic state displaying enhanced antiferromagnetic and
$d_{x^2-y2}$-wave pairing correlations. Finally, for
$|\mu|/t\geq0.1$, the system shows a $d_{x^2-y2}$-wave
superconducting phase. Moreover, if we suppose that it is possible
to associate the RG scale $\Lambda_{c}$ with an effective critical
temperature scale $T_{c}$, this could in principle provide
approximate estimates of the critical temperatures for the various
phase transitions displayed by this model obtained within our
two-loop RG scheme. Following this approach, we obtain that, for the
$d_{x^2-y2}$-wave superconducting phase, the highest effective
critical temperature scale displayed in our two-loop RG phase
diagram is given by the curve fitting value $\Lambda_{c}/4t\approx
0.004t$ (see inset), which is surprisingly in qualitative agreement
with the result $T_{c}\approx 0.023t$ obtained in the literature for
$U=4t$ within quantum cluster methods \cite{Maier}.

\section{Conclusion}

We have analyzed the 2D Hubbard model starting from
weak-to-moderate couplings by implementing a functional
generalization of the field-theoretical renormalization group
approach up to two-loop order. Our calculations here were mainly
restricted to the limit of $T\rightarrow0$, i.e. to the analysis of the
universal ground state properties and its low-lying excitations, as a function of doping. The
two-loop RG scheme was necessary in order to discuss the effect of
the momentum-resolved anomalous dimension $\eta(\textbf{p})$, which
shows up in the normal phase of the model on the corresponding low-energy
single-particle excitations. As a result, we have found
evidence for an extended antiferromagnetic insulating (AFI) ground
state in the vicinity of half-filling. This result was particularly
encouraging since it agrees with the fact that, exactly at
half-filling, this model is known to produce an AFI ground state for
any finite local interaction $U$. Our two-loop RG approach therefore
successfully reproduced this important aspect of the problem.

For a slightly higher doping regime, we have obtained that the
formation of a NFL metallic phase with a truncated Fermi surface in
the nodal regions was favored. This effect is originated by the
anisotropy of the momentum-resolved anomalous dimension
$\eta(\textbf{p})$ at this regime and it is reminiscent, to some
extent, of the nodal-antinodal dichotomy observed in the pseudogap
phase in underdoped cuprates. However, the anisotropy found here was
still very small compared with the experimental values observed in
those materials. This is undoubtedly related to our choice of
initial interaction $U=3t$. If we increase the on-site interaction
of the model, we expect that this trend will be amplified further.
Such a NFL character was observed for a finite doping range, after
which both nodal and antinodal quasiparticle weights became nonzero
even when we extrapolated our RG flow to very low RG scales. This
indicated that the resulting FS, in this new phase, was fully
reconstructed. With further doping, the system finally developed a
$d_{x^2-y^2}$-wave pairing instability in the low-energy limit. This
suggests the formation of $d_{x^2-y^2}$-wave superconducting ground
state sufficiently away from half-filling. This latter two-loop RG
result therefore adds further support to the still much-debated
existence of a $d_{x^2-y^2}$-wave superconducting phase in the 2D
Hubbard model and this is in qualitative agreement with the first
results obtained in Refs. \cite{Zanchi,Metzner,Honerkamp} within a
one-loop RG scheme.

Since our two-loop RG results have several analogies with the
observed phenomena in the high-Tc superconductors, a few remarks are
in order. These materials are known to be Mott insulators at
half-filling regime. This unambiguously implies that the
strong-coupling regime should be an important ingredient to describe
quantitatively their observed properties. The main aim of this work
however was not to achieve quantitative agreement with the
experimental results. Our objective was to show that even the
simplest version of the 2D Hubbard model is rich enough to capture
some of the main features displayed by those materials. In order to
describe quantitatively the experimental results of the high-Tc
cuprates, it is essential to choose initial couplings larger than
the ones considered in this work. Moreover, we should also take into
account the effects produced by next nearest neighbor hoppings $t'$
and $t''$. Nevertheless, it is reasonable to expect that, at least
for the interaction strengths that apply to these compounds, the
inclusion of $t'$ and $t''$ in the Hamiltonian should not alter
qualitatively the low-energy dynamics of the system. In this sense,
it is reasonable to suppose that the cuprates could be potentially
associated with a universality class which is well captured by our
model. As we have shown here, even our simple version of the 2D
Hubbard model already displays some important and nontrivial aspects
of the physics exhibited by these materials.

On the other hand, there are still some open issues which were not
addressed in this work. How the renormalization of the FS induced by
interactions affects the RG flow is still not considered in our RG
scheme. In recent years, there has been some progress in analyzing
this problem in quasi-one dimensional systems
\cite{Ledowski,Ledowski2}. However, a two-loop RG calculation of
this effect in two-dimensional systems still remains a very
difficult task and, to our knowledge, has not been calculated so
far. Another important aspect which deserves future investigation
concerns the actual importance of quantum fluctuations beyond the
two-loop RG level for this model. Since three-loop RG calculations
do not seem to be an easy prospect in the near future, we can
nevetherless learn a lot about this question by comparing one-loop
with two-loop RG calculations. One important result obtained here
was the certainty that one-loop RG calculations in general tend to
overestimate the effect of fluctuations near half-filling. As a
result, our two-loop RG theory yields a better controlled
approximation in the low-energy limit for this doping regime. In
addition, our results suggest that the pseudogap phase might be
indeed related to a strong-coupling regime in the model. However,
like any other strong-coupling problem in correlated systems, it is
still not easy to have full analytical control in such case. In
other words, there is still room for improvement before one can
claim to have a complete theory for the strong-coupling regime of
the 2D Hubbard model. However, given our encouraging results so far,
we believe our two-loop RG approach could provide a good basis for a
complementary view of this important problem starting from a
weak-to-moderate coupling perspective.

\textit{Note added} - Recently, we learned
about a related study in the context of the 2D $t-t'$ Hubbard model, Ref. \cite{Katanin2}.

\acknowledgments

One of us (HF) wants to thank the Max-Planck Institute for Solid State Research for the kind
hospitality and for financial support. Two of us (EC and AF) acknowledge support by CNPq, FINEP, and MCT
(Brazil). One of us (HF) also would like to thank Hiroyuki Yamase for a critical
reading of the manuscript and for several useful comments.


\begin{thebibliography}{99}
\bibitem{comment} Non-conserving many-body approximations in turn may lead to the violation of exact results such as Ward identities.
\end{thebibliography}
\end{document}